\documentstyle[preprint,aps]{revtex}
\tightenlines

\newcommand{\be}{\begin{equation}}
\newcommand{\ee}{\end{equation}}
\newcommand{\bea}{\begin{eqnarray}}
\newcommand{\eea}{\end{eqnarray}}

\begin{document}

\begin{titlepage}
\begin{center}
\vskip .2in
\hfill
\vbox{
    \halign{#\hfil         \cr
           hep-th/9901139 \cr
           SU-ITP-9901 \cr
           January 1999    \cr
           }  
      }   
\vskip 0.5cm
{\large \bf Conformal Mechanics and the Virasoro Algebra}\\
\vskip .2in
{\bf  J. Kumar}
\footnote{e-mail address:jkumar@leland.stanford.edu}\\
\vskip .25in
{\em
Department of Physics,
Stanford University, Stanford, California 94305 USA \\}
\vskip 1cm
\end{center}
\begin{abstract}We demonstrate that any scale-invariant mechanics of one
variable exhibits not only 0+1 conformal symmetry, but
also the symmetries of a full Virasoro algebra.  We
discuss the implications for the adS/CFT correspondence.
\vskip 0.5cm
 \end{abstract}
\end{titlepage}
\newpage

\section{Introduction}

Recently, there has been much interest in the conjectured
correspondence between theories of gravity in anti-de Sitter
background and conformal field theories living on the boundary of
$adS$ \cite{Mald} \cite{GKP} \cite{Witten}.  A particularly
interesting example of this is the $adS_{2}$/$CFT_1$ correspondence
\cite{Andy} \cite{Nakatsu} \cite{MMS}.  In particular, Strominger
\cite{Andy} pointed out that quantum gravity on $adS_2$ must be a
conformal theory on a strip, which exhibits the symmetries of the
Virasoro algebra.  This suggests that perhaps the boundary
conformal theory also exhibits the symmetries of the full Virasoro
algebra despite the fact that, naively, the boundary theory is
quantum mechanics.  This suggestion is lent credence by a recent
proposal of Gibbons and Townsend \cite{GibTown}, who suggest that
the boundary theory for $adS_2$ is given by a Calogero model of
quantum mechanics.  This model is a limit of a more general
Calogero model, which was shown to exhibit the symmetries of a
Virasoro algebra \cite{Berg}.

In fact the $SO(1,2)$ isometry group of $adS_2$, which forms the 0+1
conformal group, is given by the charge-free subalgebra of the Virasoro
algebra generated by $L_1$, $L_0$ and $L_{-1}$.  A conformal mechanics
describing the dynamics of a particle in the background of a Reisnner-
Nordstrom black hole was given in  \cite{CDKKTV}.  Motivated by \cite{Andy},
 we consider the system given in \cite{CDKKTV} to determine
if one can find generators of the full Virasoro algebra.  Not
only do we find generators of the full algebra, but we also find
(somewhat surprisingly) that given any scale-invariant Hamiltonian
of one dynamical variable, one can classically find generators of
the full Virasoro algebra (the global subalgebra of which generates
0+1 conformal symmetry).

\section{Finding the Generators}

One can write the 0+1 dimensional conformal algebra in the following
manner

\begin{eqnarray}
[D,H]=H
\qquad
[D,K]=-K
\qquad
[H,K]=2D
\end{eqnarray}
where $H$ is the Hamiltonian, $D$ is the generator of dilatations, and
$K$ is the generator of the special conformal symmetry.  By making the
identification

\begin{eqnarray}
L_0 = -\imath D
\qquad
L_{-1} = -\imath H
\qquad
L_1 = \imath K
\end{eqnarray}
we see that the $L_0$, $L_{-1}$ and $L_1$ satisfy the Virasoro
algebra

\be
[L_{m} , L_{n} ] = -\imath (m-n)L_{m+n} + A(m) \delta_{m+n}.
\ee
In \cite{CDKKTV}, we find

\begin{eqnarray}
D={1\over 2}pr
\qquad
H={p^2\over 2f}
\qquad
K=-{1\over 2}fr^2
\end{eqnarray}

\be
f={1\over 2}m[\sqrt{1+({rp\over 2mM})^2 } +1].
\ee
We take $m=1$ and take the non-relativistic massive limit, which
corresponds to $f\longrightarrow 1$ (this means that we must scale
the $r$ and $p$ by appropriate powers of the mass, giving $p$
weight $1\over 2$ and giving $r$ weight $-{1\over 2}$). We now see
if we can extend the algebra. In this case, we can rewrite part of
the envisioned Virasoro algebra as

\be
[L_{-1} , L_n ] = \imath (n+1)L_{n-1} = \imath p{\partial L_n
\over \partial r}
\ee

\be
[L_n , L_1 ] =-\imath (n-1)L_{n+1} = \imath r {\partial L_n \over
\partial p}
\ee
where we evaluate the algebra classically (ie. using Poisson brackets)
\footnote{We use the convention $[r,p]=1$}.  Using these two relations,
we find

\be
L_n = {r^2 \over n-2} {\partial^2 L_{n-1} \over \partial r
\partial p}
+{(n+1)r \over p} L_{n-1}.
\ee
Given that we have $L_0 = {1\over 2}rp$, we can use this recursive relation to
see that we must have

\be
L_n = -{\imath\over 2} r^{1+n} p^{1-n}.
\ee

Now, we must check that these $L$'s satisfy the Virasoro algebra.

\begin{eqnarray}
[L_m , L_n] &=& -{1\over 4}  [r^{1+m}p^{1-m},r^{1+n} p^{1-n}]
\nonumber\\
\qquad
&=& -{1\over 4} ( [(1+m)(1-n)-(1-m)(1+n)]r^{1+n+m} p^{1-(n+m)})
\nonumber\\
\qquad
&=& -\imath (m-n)L_{n+m}
\end{eqnarray}

We thus confirm that we in fact have the generators of the full
Virasoro algebra. One should also note that the $L$'s we have found
do not seem to obey the usual relationship $L_{n} ^\dagger =L_{-n}
$.

\section{The General Case: Scale-Invariant Hamiltonians}

We now consider the most general scale invariant Hamiltonian we can write, given that
$p$ has dimension ${1\over 2}$ and $r$ has dimension $-{1\over 2}$

\be
H={p^2 \over 2f(u)}
\ee
where $f$ is an arbitrary function of $u=rp$.  Motivated by our previous calculation
and the results of \cite{CDKKTV}, we make the following ansatz,

\be
L_n = -{\imath \over 2} r^{1+n} p^{1-n} f^n .
\ee
Note that this reduces to the case in \cite{CDKKTV} and to the free particle
case described above in the appropriate limits.
Again, we must check that the full Virasoro algebra is satisfied.
\begin{eqnarray}
[L_m , L_n] &=& -{1\over 4}  [r^{1+m}p^{1-m}f^m,r^{1+n} p^{1-n}f^n]
\nonumber\\
\qquad
&=& -\imath (m-n)L_{m+n} - {1\over 4}
[r^{2+n+m}p^{2-(n+m)}f^{n+m-1}{\partial f \over
\partial u} ] [(1+m)n\nonumber\\
&-&(1-m)n +(1-n)m - (1+n)m] \nonumber\\
\qquad
&=&-\imath (m-n)L_{m+n}
\end{eqnarray}
This shows that given any scale invariant Hamiltonian of one variable, one
can construct (classically) not only the generators of 0+1 dimensional
conformal symmetry, but also the generators of the full Virasoro algebra.

\section{The Relativistic Limit}

We will now consider the massless relativistic limit.  In this case,
we do not scale $r$ and $p$ by powers of a mass.  Thus, $r$ has weight
$-1$ and $p$ has weight 1.  We may rewrite this case in terms of our
previous analysis by making the following change of variables

\be
\tilde r = r^2
\qquad
\tilde p = {\partial L \over \partial \dot{ \tilde r}} =
p {\partial \tilde r \over \partial r} = {p\over 2r}
\ee
where we see, as expected, that $\tilde r$ has weight -1 and
$\tilde p$ has weight 1
Note that $rp = {\tilde r \tilde p
\over 2}$.
If we now choose

\be
f(rp) = rp \tilde f (\tilde r \tilde p)
\ee
then we find

\be
L_{n} = -{\imath \over 2} r^{1+n} p^{1-n} f^n
=-{\imath \over 2} r^{1+2n} p {\tilde f}^n
=-\imath {\tilde r}^{1+n} \tilde p {\tilde f}^n .
\ee
Let us set $\tilde f = 1$.
We see that $L_{-1} = -\imath \tilde p$ is then related to
the relativistic Hamiltonian $H=\tilde p$ in the expected
manner
\footnote{We may instead consider the Hamiltonian
$H=\left| p\right|$.
This may be achieved this by setting $\tilde f =1$ for positive
$p$ and $\tilde f =-1$ for negative $p$.  Note that there are
technical issues associated with the singular behavior at
$p=0$.}.
The charge-free subgroup provides a representation
of the 0+1 conformal
group as well \cite{jorjadze}.

One reason why we consider the relativisitic limit here is because
it can easily be linked to the the representation of the Virasoro algebra
in terms of diffeomorphisms of a circle.  From that point of view,
one may write the generators of the algebra as

\be
L_n = \imath e^{\imath n\theta} {\partial \over \partial \theta} .
\ee
Consider then the transformation

\be
\tilde r =e^{\imath \theta}
\qquad
L_n = \imath e^{\imath n\theta} {\partial \over \partial \theta}
=-{\tilde r}^{1+n} {\partial \over \partial \tilde r}
=-\imath {\tilde r}^{1+n} \tilde p .
\ee
Note, however, that the condition that $\tilde r$ be real amounts
to the constraint that either $\theta = C$ or $\theta = \pi + C$
where $C$ is purely imaginary.  In some sense, we can think of the
Virasoro algebra (in the relativistic free particle limit, at least)
as arising from the diffeomorphisms of a circle, analytically continued
to a real line.

\section{Computing the Central Charge}

The analysis of the previous section also provides insight into
the quantization of the Virasoro algebra.  The transition
$-\imath {\partial \over \partial \tilde r} \longrightarrow
\tilde p$ is essentially the quantization of the system.  The
transformation of the previous section then seems to give us
an ordering for the quantum generators.  In particular, we
make the ansatz

\be
L_n = -\imath {\tilde r}^{1+n} \tilde p
\ee
as a quantum operator.
We then examine the full quantum algebra, and find

\begin{eqnarray}
[L_n , L_m] &=& -[{\tilde r}^{1+n} \tilde p ,{\tilde r}^{1+m}
\tilde p] \nonumber\\
\qquad
&=& -{\tilde r}^{1+n} [\tilde p, {\tilde r}^{1+m}
\tilde p] - [{\tilde r}^{1+n}, {\tilde r}^{1+m}
\tilde p] \tilde p  \nonumber\\
\qquad
&=& -{\tilde r}^{1+n} [\tilde p, {\tilde r}^{1+m}
]\tilde p - {\tilde r}^{1+m}[{\tilde r}^{1+n},
\tilde p] \tilde p \nonumber\\
\qquad
&=& -\imath (n-m) {\tilde r}^{1+n+m} \tilde p
=(n-m) L_{n+m} .
\end{eqnarray}
In the case of the free relativistic particle, we have exhibited
generators which satisfy the full quantum Virasoro algebra with
no central charge.

\section{Multiplets of States}

One might expect to be able to find the states organized into
highest weight representations of the Virasoro algebra.  In fact,
this is somewhat  difficult.  In 1+1 conformal field theory,
$L_0$ is the Hamiltonian and is bounded from below. Since all of
the other $L$'s will shift the eigenvalue of $L_0$, one finds that
the eigenstate of $L_0$ corresponding to its minimum eigenvalue
must be annihilated by the half of the $L$'s which are lowering
operators.  It is upon this eigenstate that one builds multiplets
of the algebra. In our case, however, $L_0$ is the generator of
dilatations, and its eigenvalues are not bounded. Instead, the
Hamiltonian is given (up to constant factors) by $L_{-1}$.
In order to find an equivalent
construction, we must find linear combinations of the $L$'s such
that

\be
[H,\sum_{n}{a_n L_n} ] = C \sum_{n} a_n L_n .
\ee
These operators will then be the raising and lowering operators of
the representation.  Using the commutation relations, one can easily
see that the raising and lowering operators will be of the form

\be
e^{\imath kr} p .
\ee
One could see this just as well by remembering that since $H=p$,
the raising and lowering operators are just operators which
shift the momentum.  The only state which is annihilated by
these operators is the zero momentum state.  But since it
is annihilated by all of these operators, one cannot build
the multiplet upon it.  In that sense, there is no highest
weight representation.

\section{Hermiticity and Global Time}

We have noted that, for the constructions provided in the previous
sections, one does not have the condition $L_n ^{\dagger} =
L_{-n}$. This seems unusual given that, in some cases, the
scale-invariant mechanics actually describes the motion of a
particle in the background of $adS_2$.  The $L_n$'s of the bulk are
the Fourier modes of the bulk stress-energy tensor, and as such must
obey the hermiticity condition.  One might ask why the same does
not appear to be true for the particle.  The answer is related to
the choice of time.

Both the free non-relativistic particle and free relativistic
particle are limits of the Hamiltonian found in \cite{CDKKTV}. That
Hamiltonian was derived from the Born-Infeld action for the radial
excitations of a particle in the background of $adS_2 \times S^2$
given in Poincar\'e coordinates.  These coordinates do not cover
the entire anti-de Sitter space.  One can write these coordinates in
such a way that the space they cover is a half-plane which is conformally
flat.

\begin{eqnarray}
ds^2 = {1\over u^2 }(-dt^2 + du^2)
\qquad
u>0
\end{eqnarray}
If one thinks of these coordinates as parametizing the upper half of the
complex plane, then time translation is the same as translation along
the real axis.  Since the $L_n$'s generate transformations of the upper
half-plane, one can see that $L_{-1}$ is the generator of translation along
the real axis.  Therefore, it is reasonable to associate the Hamiltonian with
$L_{-1}$.  One then easily sees that one can identify $L_{-1,0,1}$ with
$H$,$D$ and $K$ in such a way that the Virasoro algebra and 0+1 conformal
algebra both close.

But this is not the usual situation with which we are familiar from
1+1 conformal field theory.  In that case, we generally find that
the Hamiltonian is given by $L_0$, and that $L_n ^\dagger =
L_{-n}$. In order to find similar relations, one must choose the
time coordinate such that time translation is generated by $L_0$.
Global time coordinates \footnote{We will actually parameterize
$CadS_2$, the covering space of $adS_2$.} are well suited for this
purpose.  These coordinates may be written as

\be
ds^2 = -(1+a^2 r^2) dt^2 + (1+a^2 r^2 )^{-1} dr^2 .
\ee
After the transformation
\begin{eqnarray}
tan(\sigma) = ar
\qquad
\rho = e^{at}
\end{eqnarray}
and a Wick rotation, one finds the metric
\begin{eqnarray}
ds^2 = {1\over a^2 \rho^2 cos^2 (\sigma) } (d\rho^2 + \rho^2 d\sigma^2)
\qquad
-{\pi \over 2}<\sigma <{\pi \over 2} .
\end{eqnarray}
It is clear that time translation now corresponds to scaling of
the half-plane, which is generated by $L_0$.

Let us now write the action of a particle in an
anti-de Sitter space background given in global time coordinates.

\be
L = -m\sqrt{(1+a^2 r^2 ) - (1+ a^2 r^2 )^{-1} {\dot r}^2}
\ee
The Hamiltonian associated with this action is

\be
H=(1+a^2 r^2)^{1\over 2} (m^2 + p^2 + a^2 r^2 p^2 )^{1\over 2} .
\ee

In the large radius, relativisitic limit the Hamiltonian reduces to
$H=p$ (as one would expect for a relativisitic particle in nearly flat
space).  Following the logic used above, we set $L_0 = H = p$.

We must now attempt to find the generators of the rest of the
Virasoro algebra.  Note that we should not expect $L_{-1,0,1}$ to
be expressed in terms of $H$, $D$ and $K$ (which is the formulation
appropriate to Poincar\'e coordinates).  If $L_0 = H$, there is
no way to write $L_1$ or $L_{-1}$ in terms of $H$, $K$ or $D$ such
that they close under the Virasoro algebra.  But one quickly sees
that in order for $[L_m , L_0 ] = -\imath mL_m $ to hold
(classically), $L_m$ must be of the form

\be
L_m = g_m (p) e^{-\imath mr}  .
\ee
We make the ansatz

\be
L_m = e^{-\imath mr} p
\ee
and can readily verify that this choice satisfies the classical Virasoro
algebra under Poisson brackets.  We also note that $L_n ^{\dagger} = L_{-n}$
(classically).  This is exactly what we hoped to find in order to match our
intuition from 1+1 CFT.

The next question is whether or not the above system can be quantized.  Consider the
ansatz
\be
L_m = e^{-{\imath m \over 2} r} p e^{-{\imath m \over 2} r}
\ee
where $L_m$ is now a quantum operator.  It is clear that
$L_m ^{\dagger} = L_{-m}$.  One also sees that

\begin{eqnarray}
[L_n , L_m] &=& e^{-{\imath n \over 2} r} [p , e^{-{\imath m \over 2} r}] p
e^{-{\imath m \over 2} r} e^{-{\imath n \over 2} r}
+e^{-{\imath n \over 2} r} e^{-{\imath m \over 2} r} p
[p,e^{-{\imath m \over 2} r}] e^{-{\imath n \over 2} r}
\nonumber\\ \qquad
&+&e^{-{\imath n \over 2} r} p e^{-{\imath m \over 2} r}
[e^{-{\imath n \over 2} r},p]e^{-{\imath m \over 2} r}
+e^{-{\imath m \over 2} r}[e^{-{\imath n \over 2} r},p]
e^{-{\imath m \over 2} r}pe^{-{\imath n \over 2} r}
\nonumber\\ \qquad
&=& (n-m)L_{n+m} + {n\over 2} e^{-{\imath n \over 2} r}
[p,e^{-{\imath m \over 2} r}]e^{-{\imath n \over 2} r}
e^{-{\imath m \over 2} r}
+ {n\over 2} e^{-{\imath m \over 2} r}e^{-{\imath n \over 2} r}
[e^{-{\imath m \over 2} r},p]e^{-{\imath n \over 2} r}
\nonumber\\ \qquad
&=& (n-m) L_{n+m} .
\end{eqnarray}
Therefore we can conclude that for the free relativistic particle
in global time, the Virasoro algebra satisfies hermiticity and has
no central charge.

One may also consider the limit of small $adS$ radius and small $m$.
In that case, we find
\be
L_0 = H=pr^2 .
\ee
Take the ansatz
\be
L_m = pr^2 e^{\imath m \over r} .
\ee
Again, we can verify that the Virasoro algebra does indeed close
under Poisson brackets.  To quantize this system, we make the ansatz
\be
L_m = e^{\imath m \over 2r} rpr e^{\imath m \over 2r}
\ee
as a quantum operator.  We then find that the quantum Virasoro algebra
closes with no central charge.

\section{Conclusion}

It is interesting to note that the $adS/CFT$ correspondence served only as
a motivational tool in the above derivation.  The conformal mechanics
described above need not be related to a theory on anti-de Sitter space.
It is worthwhile to further investigate the relationship
between conformal mechanics and gravity theories on $adS_2$ \cite{CKV}
\cite{Gomis}.

Additionally, the arguments given above only apply to the mechanics of a
single variable.
It would be interesting to conjecture that any conformal
mechanics of any arbitrary number of degrees of freedom contains the symmetries
of the Virasoro algebra.  If true, this would more closely cement the relationship
found in \cite{Andy} between conformal quantum mechanics and 1+1
conformal field theory.

Finally, it would very interesting to completely understand the quantization
of these conformal mechanics models.  More concretely, one would like to know
if there is a way to determine the central charge and normal-ordering conventions
for arbitary scale-invariant Hamiltonians.  Thus far, all of the systems which
we have been able to quantize could be thought of as describing a limit of a
particle in the background of $adS_2$.  Since the bulk theory is a
two-dimensional quantum gravity
theory, the central charge of the bulk Virasoro algebra should vanish (when all
ghosts are included).  One might speculate that this is the reason why the
central charge of the quantum mechanics has vanished in all cases we have found.
One would like to know if a non-vanishing central charge could perhaps be found
for quantized systems which did not represent a particle in the background of
$adS_2$.

\acknowledgements

We gratefully acknowledge S. Cullen, J. Gomis, E. Halyo, R.
Kallosh, S. Shenker, E. Silverstein, Y. Song, N. Toumbas and
A. Van Proeyen for
useful discussions. We especially thank A. Strominger and L.
Susskind for very helpful comments.  This work is supported by the
Department of Defense, NDSEG Fellowship Program and by NSF grant
PHY-9870115.

\end{document}